\documentclass[conference]{IEEEtran}
\IEEEoverridecommandlockouts
\usepackage[utf8]{inputenc}
\usepackage{cite}
\usepackage{amsmath,amssymb,amsfonts}
\usepackage{algorithmic}
\usepackage{graphicx}
\usepackage{textcomp}
\usepackage{xcolor}
\usepackage[hidelinks]{hyperref}
\usepackage{fancyhdr}
\def\BibTeX{{\rm B\kern-.05em{\sc i\kern-.025em b}\kern-.08em
    T\kern-.1667em\lower.7ex\hbox{E}\kern-.125emX}}
\begin{document}

\title{Bytes to Threats: Byte-Sequence Transformer For Malware Detection}

\author{\IEEEauthorblockN{1\textsuperscript{st} Abdelrahman Yassin Eljamal}
%\author{\IEEEauthorblockN{Abdelrahman Yassin}
\IEEEauthorblockA{\textit{Department of Computer Science} \\
\textit{University of Rhode Island}\\
Kingston, USA \\
abdelrahman.eljamal@uri.edu}
\and
\IEEEauthorblockN{2\textsuperscript{nd} Abdeltawab Hendawi}
%\IEEEauthorblockN{Abdeltawab Hendawi}
\IEEEauthorblockA{\textit{Department of Computer Science} \\
\textit{University of Rhode Island}\\
Kingston, USA \\
hendawi@uri.edu}
}

\maketitle
\begingroup
\renewcommand\thefootnote{}
\footnotetext{\footnotesize © 2026 IEEE. Personal use of this material is permitted. Permission from IEEE must be obtained for all other uses, in any current or future media, including reprinting/republishing this material for advertising or promotional purposes, creating new collective works, for resale or redistribution to servers or lists, or reuse of any copyrighted component of this work in other works. DOI: 10.1109/ICAD69378.2026.11608884}
\endgroup

\begin{abstract}
Existing malware detection approaches heavily rely on handcrafted features derived from static or dynamic analysis of the executable files. These methods require specialized domain expertise, dedicated analysis tools, and time-consuming analysis, which limits their scalability and robustness in the face of the increasing complexity and diversity of modern malware. Consequently, there is a growing interest in approaches that operate directly on raw bytes, which contain rich semantic and structural information. However, capturing long-range dependencies within executable files remains challenging for conventional static and CNN-based methods. This paper presents \textbf{\mbox{MalwareFormer}}, a transformer-based, byte-level malware detection framework that operates directly on raw bytes, without requiring domain knowledge or feature extraction. \textbf{\mbox{MalwareFormer}} achieves an accuracy of 
\textbf{98.35\%}, a balanced accuracy of \textbf{98.32\%}, a macro-F1 of 
\textbf{98.32\%}, and a false positive rate of \textbf{1.94\%}, outperforming both 
baselines across all threshold-dependent classification metrics and demonstrating that 
transformers provide a powerful and scalable approach for byte-level malware detection.
\end{abstract}


\section{Introduction}
The rapid growth of malware variants poses a significant and escalating economic threat. Recent estimates project that the global cost of cybercrime will reach approximately \$10.5 trillion annually by 2025, underscoring its scale as one of the largest economic threats worldwide. At the organizational level, the average cost of a single data breach is estimated at \$4.44 million globally. Ransomware and phishing attacks remain dominant, accounting for approximately 44\% and 16\% of reported breaches, respectively~\cite{deepstrike-2025}.
In the past year alone, cyber incidents led to substantial financial losses, including an estimated \$3.09\, billion impact on UnitedHealth and a \$50\, million ransom demand targeting CDK Global~\cite{sygnia-2025}. According to Kaspersky~\cite{kaspersky-2024}, their detection systems identified an average of 467{,}000 malicious files per day in 2024, with Trojan-based attacks showing a notable increase compared to 2023. Despite continuous advancements, anti-virus technologies are struggling to keep pace with the rapid evolution and sophistication of modern malware, highlighting the need for more adaptive detection methods.

Existing machine learning approaches rely heavily on handcrafted features extracted from executable files, including header information, imported libraries, and opcode statistics~\cite{M2023ACS, Martins2025SemanticMC}. While effective, these methods require significant domain expertise and extensive effort to extract such features. For instance, dynamic analysis captures behavioral and runtime features by executing malware in a controlled environment, such as a virtual machine. This allows the collection of system call traces, registry changes, file and network activity, and process behaviors, but the approach is slow, resource-intensive, and difficult to scale. Static analysis, in contrast, operates directly on the file without execution, making it faster and less resource-demanding. However, its effectiveness remains limited by the handcrafted features and specialized tools required for tasks such as extracting PE header information, imported libraries, opcode sequences, strings, and metadata, as well as performing reverse engineering and parsing of embedded resources, all of which necessitate expert knowledge.

Deep learning approaches have emerged as a promising alternative by enabling direct processing of raw bytes without feature engineering. CNN-based models such as MalConv~\cite{Raff2017MalwareDB} demonstrated the feasibility of learning patterns from entire executable files; however, their reliance on local convolutional filters limits their ability to capture long-range dependencies inherent in executable files. Transformer architectures, with their global self-attention mechanism, offer a better framework for modeling byte sequences.

To address these limitations, this paper proposes \textbf{\mbox{MalwareFormer}}, a transformer-based, byte-level malware detection framework designed to operate directly on raw bytes without requiring hand-crafted features or domain knowledge.
At its core, \textbf{\mbox{MalwareFormer}} leverages a transformer model, ByteFormer~\cite{Horton2023BytesAA}, which is fine-tuned within our framework to efficiently capture long-range dependencies in malware byte sequences. We conducted extensive experiments to evaluate the model’s effectiveness, comparing training from scratch with fine-tuning and analyzing their impact on convergence speed and final performance. We further examine how varying input sequence lengths influence model accuracy and stability. 
To contextualize its performance, we compare \textbf{\mbox{MalwareFormer}} against 
two strong CNN-based baselines, MalConv2~\cite{Raff_Fleshman_Zak_Anderson_Filar_McLean_2021} 
and AvastNet~\cite{Krcl2018DeepCM}, using a 50K-byte input size. 
\textbf{\mbox{MalwareFormer}} achieves an accuracy of \textbf{98.35\%}, a balanced 
accuracy of \textbf{98.32\%}, a macro-F1 of \textbf{98.32\%}, and a false positive 
rate of \textbf{1.94\%}, outperforming both baselines across all threshold-dependent 
classification metrics. MalConv attains 95.71\% accuracy, a macro-F1 of 95.64\%, and 
a false positive rate of 3.56\%, while AvastNet achieves 97.98\% accuracy, a macro-F1 
of 97.94\%, and a false positive rate of 2.46\%. Notably, while AvastNet attains a 
marginally higher ROC-AUC (99.68\% vs.\ 99.61\%), MalwareFormer yields a better 
trade-off between detection rate and false alarm rate at the operating threshold used 
in practice. These results demonstrate that transformer-based byte modeling enables 
\textbf{\mbox{MalwareFormer}} to capture long-range structural dependencies within 
executables that CNN-based approaches struggle to model, translating into more reliable detection under realistic deployment conditions.

This paper investigates the capabilities of transformer-based, byte-level malware detection by addressing the following research questions. 
\textbf{RQ1:} \textit{How effective is a transformer-based model compared to established CNN-based baselines in classifying malware?} 
\textbf{RQ2:} \textit{How does fine-tuning pretrained representations impact training convergence and detection performance?} 
\textbf{RQ3:} \textit{Can accurate detection be achieved without processing the entire executable file?} 
\textbf{RQ4:} \textit{How does entropy affect classification accuracy, and what asymmetric failure modes emerge in high-entropy regions?}

Our key contributions are summarized as follows:
\begin{itemize}
    \item We present \textbf{\mbox{MalwareFormer}}, a byte-level transformer framework 
    for malware detection that incorporates dataset integrity verification, label 
    consistency validation, byte-sequence extraction, and model fine-tuning, achieving 
    \textbf{98.35\%} accuracy and a false positive rate of \textbf{1.94\%} without requiring handcrafted features or domain expertise.
    \item Investigate the transferability of pretrained ByteFormer representations to the malware domain by fine-tuning the model and analyzing their impact on training convergence and detection performance across varying input sequence lengths. To the best of our knowledge, this is the first work to utilize the ByteFormer model for malware detection.
    \item Compare this transformer-based model with established CNN-based baselines, providing empirical evidence of its effectiveness for static malware detection.
    \item We conduct an entropy-stratified analysis of model predictions, revealing an 
    asymmetric failure mode in high-entropy regions (entropy 7--8) where FPR reaches 
    14.24\% while FNR drops to 0.08\%, exposing a fundamental limitation of byte-level 
    detection in entropy-ambiguous regimes and motivating future entropy-aware approaches.
\end{itemize}

The rest of our paper is organized as follows. Section~\ref{sec:related_work} reviews related work. Section~\ref{sec:method} presents the \textbf{\mbox{MalwareFormer}} framework, detailing its transformer-based, byte-level architecture and the dataset employed in our experiments. Section~\ref{sec:experiment_setup} provides the experimental setup. Section~\ref{sec:result} presents the experimental results. Section~\ref{sec:discussion} provides an analysis and discussion of the results. Finally, Section~\ref{sec:conclusion} concludes our work.

\section{Related Work}
\label{sec:related_work} 
This section reviews prior research on malware detection using raw bytes as input, beginning with convolutional neural network (CNN)–based approaches and progressing to more recent Transformer‑based architectures.

\subsection{CNN Based Models}
Early work demonstrated that deep learning models can effectively analyze malware directly from raw executable bytes, eliminating the reliance on manual feature engineering. Reference~\cite{Raff2017MalwareDB} pioneered this direction with MalConv, a convolutional neural network designed specifically for Windows PE files. MalConv processes executable files as byte sequences using a gated convolutional layer followed by global max pooling and a fully connected classifier. Their results showed that meaningful representations can be learned directly from raw bytes, capturing global structural patterns present within the entire executable file.
Building upon this idea, reference~\cite{Krcl2018DeepCM} investigated deeper convolutional architectures trained exclusively on raw executable files. Their study confirmed that CNNs can generalize well without hand-crafted input features and systematically analyzed architectural factors that influence detection performance. These findings strengthened the evidence that binary‑level features can be learned directly from byte sequences.
A major challenge in byte-level malware analysis is the extreme input length, which can exceed several million bytes. This limitation was addressed by proposing memory‑efficient sequence classification methods with constant memory complexity, enabling processing of very long byte sequences~\cite{Raff_Fleshman_Zak_Anderson_Filar_McLean_2021}. Their approach increased the scalability of raw‑byte malware detection and facilitated practical large-scale deployment. Other works focus on non-executable documents such as PDF files. Reference~\cite{Jeong2019MalwareDO} models PDF files as raw byte streams and applies convolutional neural networks to learn discriminative byte-level representations without relying on explicit file parsing. Their approach captures spatial and structural patterns within byte sequences to identify malicious content embedded in non-executable documents. In contrast, byte-stream–based detection extended to the file level by aggregating stream-level predictions across an entire document~\cite{Jeong2023FilelevelMD}. This aggregation strategy enables more robust file-level malware detection, improving detection performance while maintaining resilience against obfuscation techniques in document-based malware.

Hybrid approaches have also been explored. In~\cite{Le2018DeepLA}, they combined CNN layers (for local feature extraction) with LSTM layers (for long‑range dependencies), demonstrating improved performance by modeling both spatial and sequential patterns in byte streams.

\subsection{Transformer Based Models}
Since the introduction of the Transformer architecture \cite{Vaswani2017AttentionIA}, Transformer-based models have achieved state-of-the-art performance across a wide range of domains beyond natural language processing, including malware detection. Reference~\cite{Kunwar2024SoKLT} provided a comprehensive survey of Transformer-based approaches for malware classification, noting that the majority of existing methods rely on engineered or extracted features rather than operating directly on raw binary data.

Several studies apply Transformers to sequential representations of program content, particularly assembly code. Works such as \cite{Li2023GenTALGD, Li2019IMADIM, Demirci2022StaticMD} model assembly instructions as token sequences, enabling Transformers to capture long-range dependencies and instruction-level semantics for malware detection.

Beyond linear sequences, In~\cite{He2024AdaTransFA}, the author introduced a Transformer architecture designed to model function-call graphs, generating graph-based embeddings that capture semantic relationships in program behavior. This approach demonstrates the flexibility of Transformers in handling structured representations beyond flat sequences.

Other researchers have explored transforming binary data into image-like representations suitable for vision Transformers. Seneviratne \emph{et al.}~\cite{Seneviratne2022SelfSupervisedVT} and subsequent works~\cite{Ravi2023ViT4MalLV, Park2022AVT} demonstrate that such visual encodings can reveal spatial patterns in malware binaries that are difficult to capture using traditional static analysis.

In addition to code and binary representations, Transformer-based models have been applied to higher-level feature types, including manifest information \cite{Rahali2021MalBERTMD, Fan2021HeterogeneousTG, Rahali2023MalBERTv2CA} and network behavior data \cite{Wangwang2021NetworkTO, Ullah2022ExplainableMD, Trizna2023NebulaSF}. These approaches further demonstrate the versatility of Transformers in modeling heterogeneous sources of malware-related information.

\subsubsection{Byte Only Transformer Models}
Despite substantial progress, most existing Transformer‑based malware detection methods continue to rely on extracted or handcrafted features (assembly instructions, graphs, images, etc.). Reference~\cite{Kim2025MalwareDU} introduced ByteEnc, a byte-level pre-trained Transformer encoder designed for non-executable document formats such as PDF, HWP, and MS Office files. The work proposed two novel pre-training objectives, Real-or-Fake Prediction (RFP) and Same-Sequence Prediction (SSP), combined with masked language modeling (MLM) and next-sentence prediction (NSP). It further demonstrated that relative positional encodings and byte-level unigram tokenization improve performance on byte-level understanding tasks.

Our work focuses exclusively on raw byte input, employing a transformer-based architecture within the proposed \textbf{\mbox{MalwareFormer}} framework to capture long-range dependencies in executable files.

\section{Methodology}
\label{sec:method}
\begin{figure}[htbp]
    \centering
    \includegraphics[height=4.3cm, width=0.48\textwidth]{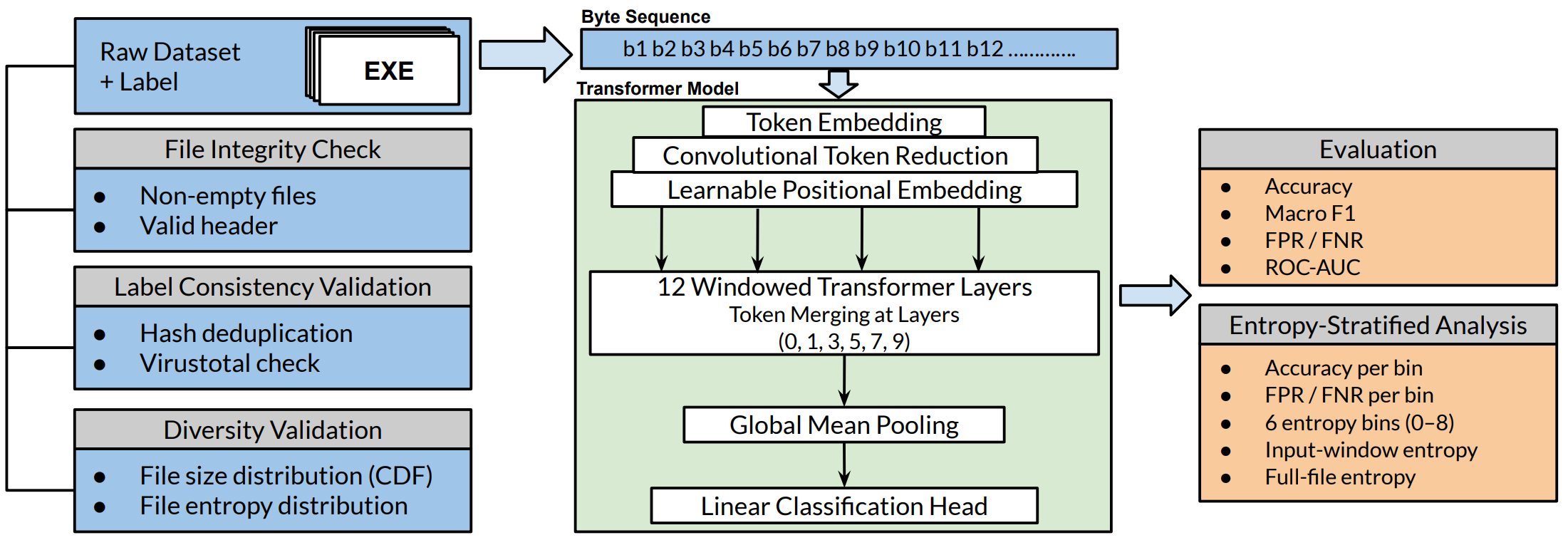}
    \caption{MalwareFormer Framework.}
    \label{fig:MalwareFormer}
\end{figure}

Fig.~\ref{fig:MalwareFormer} illustrates our proposed framework, 
\textbf{\mbox{MalwareFormer}}. The process begins with an integrity check of the 
dataset files, followed by label consistency validation through cross-verification 
of samples using VirusTotal~\cite{VirusTotal}. A diversity check is then conducted 
to ensure variability among samples in terms of file size and entropy. In the second 
phase, raw byte sequences are fed into the model described in the Model Overview 
subsection (\ref{model_overview}). Finally, the model is evaluated using accuracy, 
balanced accuracy, macro-F1, false positive rate (FPR), false negative rate (FNR), 
and ROC-AUC. We further conduct an entropy-stratified analysis of model predictions 
across six entropy bins to identify entropy-induced false positive bias and 
characterize regimes where byte-level detection is inherently limited.

\subsection{Model Overview}
\label{model_overview}
We follow the Byteformer model \cite{Horton2023BytesAA}. It is a transformer-based encoder model designed to capture contextual representations directly from byte-only input sequences without requiring any modality-specific preprocessing. The model consists of a learnable positional embedding table, an initial convolutional token-reduction module that downsamples the raw byte stream, and a stack of 12 windowed transformer layers. Within these 12 layers, 6 token merging operations are distributed at layers (0, 1, 3, 5, 7, 9) to hierarchically reduce the sequence length and aggregate local byte-level patterns. Each transformer layer applies multi-head self-attention with windowed attention, followed by a feedforward network. After the final transformer block, a global mean pooling operation is applied to the token representations before feeding them into a linear classification head.

\subsection{Dataset}
For this task, we used ''PE Malware Machine Learning Dataset''~\cite{practicalsecurityanalyticsMalwareMachine}, which is collected from multiple public repositories, including VirusShare, MalShare, and TheZoo, and contains 201,549 executable files, 86,812 benign, and 114,737 malicious files.

\begin{figure}[htbp]
    \centering
    \includegraphics[height=6cm,width=0.48\textwidth]{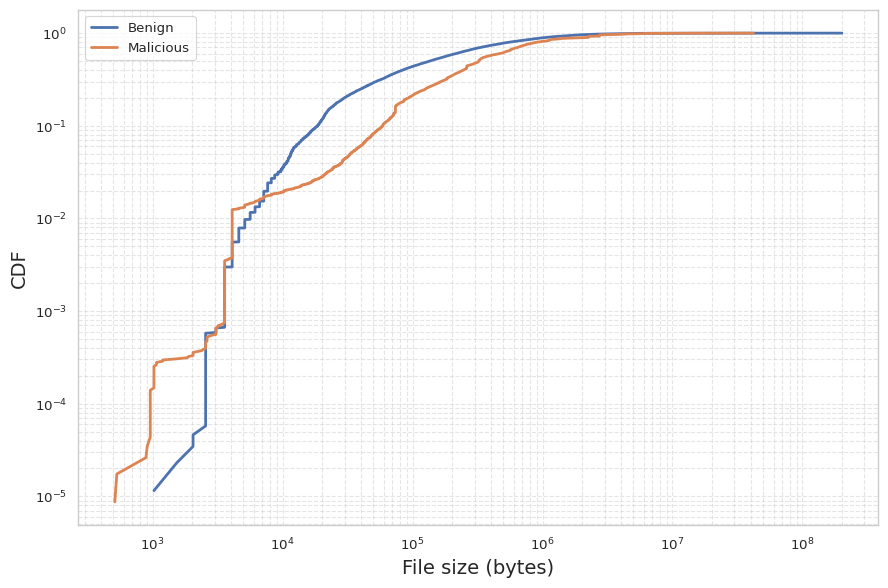}
    \caption{Cumulative Distribution of File Sizes}
    \label{fig:cdf_datasetA}
\end{figure}

Fig.~\ref{fig:cdf_datasetA} illustrates the cumulative distribution function (CDF) of file sizes for benign and malicious executable files. The two curves are highly overlapped, indicating that benign and malicious samples share very similar size distributions overall. Benign executable files predominantly start at file sizes above $10^3$ bytes, while malicious samples include smaller binaries, with the distribution extending below $10^3$ bytes. In particular, malware files exhibit finer-grained steps at smaller sizes, with approximately five distinct steps down to 512 bytes and a non-negligible cumulative probability above $10^{-3}$. Despite this difference in the lower-size regime, the two curves converge quickly as file size increases, remaining close across the majority of the distribution. This substantial overlap suggests that file size alone is insufficient to reliably distinguish benign and malicious executable files, motivating the need for content-aware byte-level modeling approaches.

\textbf{Entropy} is employed as a descriptive metric to characterize the statistical properties of the dataset. File entropy is computed using Shannon entropy with probabilities estimated via the maximum likelihood estimator (MLE) over byte-level distributions, which provides insight into the degree of randomness present in benign and malicious files; therefore, entropy is used as an indicator for high randomness, packing, compression, and encryption in executable files \cite{deGaspari2021ReliableDO,4140989}.

\begin{figure}[htbp]
\centerline{\includegraphics[height=6cm,width=0.48\textwidth]{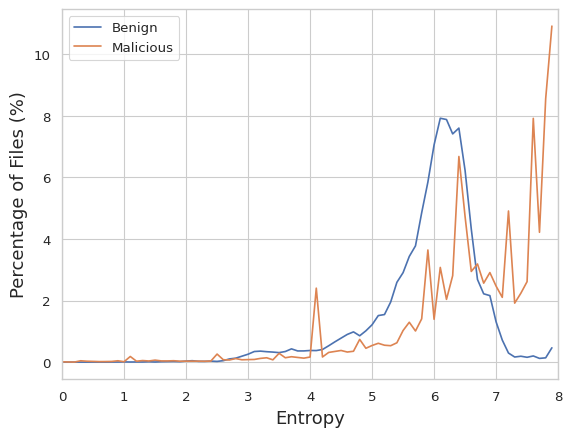}}
\caption{Percentage of Files as a Function of Entropy (0–8)}
\label{fig:entropy_datasetA}
\end{figure}

Fig.~\ref{fig:entropy_datasetA} shows a peak in the benign distribution around entropy of 5.8 to 6.3, which corresponds to typical compiled executable files that contain a mixture of machine code, static data, and limited compression.
In contrast, malware distribution is more dispersed, indicating greater structural variability in malware binaries. Another notable feature is the presence of localized spikes in the malicious entropy curve, especially in the high-entropy tail. These spikes suggest that malware does not form a smooth statistical population but instead consists of multiple sub-populations, likely corresponding to different malware families or packing strategies. Each family may rely on a specific packer or encryption scheme, resulting in characteristic entropy values. 

\section{Experimental Setup}
\label{sec:experiment_setup}
We conducted multiple experiments to study its behavior under different training strategies, including fine-tuning from a model pre-trained on image bytes and training from scratch on executable files. In addition, we evaluated the impact of different input lengths by feeding the model the first 50K, 100K, 150K, and 200K bytes of each executable. This analysis is motivated by the large variation in file sizes within our dataset, which ranges from approximately 500 bytes to more than 2 million bytes for larger files.

All experiments were conducted within the CoreNet framework \cite{apple-no-date}, which natively supports ByteFormer. To ensure a fair and reproducible comparison, we integrated the additional baseline models (MalConv and AvastNet) into the same framework and trained each model using the hyperparameter configuration reported in its original publication whenever possible. For consistency, all models were trained for 30 epochs.

For ByteFormer, training was performed using the AdamW optimizer with a weight decay of 0.05, excluding batch normalization layers and bias parameters from weight decay. The optimizer parameters were set to $\beta_1=0.9$ and $\beta_2=0.999$. A cosine learning rate scheduler was used with a linear warmup of 1,000 iterations, using an initial learning rate of $1\times10^{-6}$, a peak learning rate of $3\times10^{-5}$, and a minimum learning rate of $1\times10^{-6}$. The scheduler was configured to update on an epoch basis.

For MalConv and AvastNet, we followed the training configuration reported by the original authors. Both models were trained with Adam optimizer ($\beta_1=0.9$, $\beta_2=0.999$) with a fixed learning rate of $1\times10^{-3}$ and no weight decay. Training was conducted for 30 epochs with a batch size of 128. The input size was fixed at 50,000 bytes, and the cross-entropy loss was used for binary classification.

\section{Results}
\label{sec:result}

\begin{figure}[htbp]
    \centering{\includegraphics[height=6cm,width=0.48\textwidth]{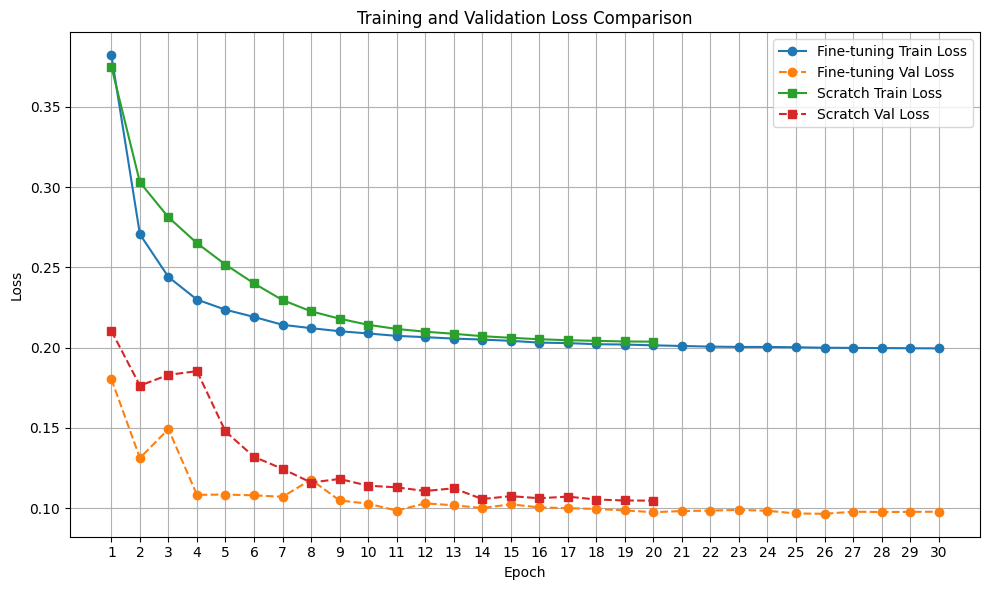}}
    \caption{Training Loss Across Epochs: Fine-Tuning vs. Training from Scratch}
    \label{fig:byteformer_loss_convergence}
\end{figure}

Fig.~\ref{fig:byteformer_loss_convergence} compares the training and validation loss curves for fine-tuning and training \textbf{\mbox{MalwareFormer}} from scratch. The fine-tuned model, which uses pre-trained weights, converges significantly faster, with the validation loss decreasing rapidly and falling below 0.1 by approximately epoch 11 of the 30-epoch training schedule. In contrast, the model trained from scratch over 20 epochs maintains a higher validation loss throughout the same period.

Although the fine-tuned model starts with a higher training loss in the first epoch, it drops sharply after the initial epochs, reaching a lower training loss than the model trained from scratch by the second epoch. These observations highlight the benefits of leveraging pre-trained representations to accelerate convergence and improve model stability.

\subsection{Effect of Input Sequence Length}
Table~\ref{tab:byteformer_results} reports the effect of input sequence length on classification performance for both fine-tuning and training from scratch. For the fine-tuned model, increasing the input size from 50K to 100K bytes results in consistent improvements in both accuracy and AUC. Further increasing the input size to 150K or 200K bytes yields only marginal gains.

Overall, these results indicate that \textbf{\mbox{MalwareFormer}} achieves stable performance even with truncated byte sequences, while larger input sizes can provide modest additional benefits.

\begin{table}[htbp]
\centering
\caption{Model results with different input sizes.} 
\label{tab:byteformer_results}
\resizebox{\columnwidth}{!}{
\begin{tabular}{|p{3cm}|p{3cm}|p{3cm}|}
\hline
\textbf{Input Size} & \textbf{Acc.\ (\%)} & \textbf{ROC-AUC.\ (\%)} \\
\hline
\multicolumn{3}{|c|}{Fine-Tuning} \\
\hline
50K & 98.13 & 99.17 \\
\hline
100K & 98.45 & 99.21 \\
\hline
150K & 98.35 & 99.45 \\
\hline 
200K & 98.48 & 99.42\\
\hline
\end{tabular}
}
\end{table}

\subsection{Comparison with CNN-Based Malware Detectors}
To evaluate the effectiveness of MalwareFormer, we compare its performance against 
two widely used CNN-based malware detection models, MalConv~\cite{Raff_Fleshman_Zak_Anderson_Filar_McLean_2021} 
and AvastNet~\cite{Krcl2018DeepCM}, using a 50K-byte input size across all models. 
As shown in Table~\ref{tab:comparison_results}, \textbf{\mbox{MalwareFormer}} achieves 
the highest overall classification performance, with an accuracy of \textbf{98.35\%}, 
a balanced accuracy of \textbf{98.32\%}, a macro-F1 of \textbf{98.32\%}, and a false 
positive rate (FPR) of \textbf{1.94\%}, the lowest among all evaluated models. 
Although AvastNet attains a marginally higher ROC-AUC (99.68\% vs.\ 99.61\%), 
MalwareFormer consistently outperforms both CNN-based baselines across all 
threshold-dependent metrics, which are more directly relevant to practical deployment 
where false alarm rates and missed detections carry real operational costs.

On the shared test set of 30,233 samples (17,211 malware, 13,022 benign), 
MalwareFormer correctly detects 16,966 malware samples while missing only 245 
(FNR: 1.42\%), and correctly classifies 12,769 benign files with just 253 false 
alarms (FPR: 1.94\%). Compared to AvastNet, this represents 46 fewer missed malware 
detections and 67 fewer false positives on benign files, demonstrating improved 
reliability under realistic class distributions.

\begin{table}[htbp]
\centering
\caption{Performance comparison of MalwareFormer, MalConv, and AvastNet 
         (50K-byte input). FPR = false positive rate; FNR = false negative rate.
         Best results per metric in \textbf{bold}.}
\label{tab:comparison_results}
\renewcommand{\arraystretch}{1.5}
\resizebox{\columnwidth}{!}{%
\begin{tabular}{|l|c|c|c|c|c|c|}
\hline
\textbf{Model} & \textbf{Acc.\ (\%)} & \textbf{Bal.\ Acc.\ (\%)} & 
\textbf{Macro F1 (\%)} & \textbf{FPR (\%)} & \textbf{FNR (\%)} & 
\textbf{ROC-AUC (\%)} \\
\hline
MalwareFormer & \textbf{98.35} & \textbf{98.32} & \textbf{98.32} & \textbf{1.94} & \textbf{1.42} & 99.61 \\
\hline
AvastNet      & 97.98 & 97.93 & 97.94 & 2.46 & 1.69 & \textbf{99.68} \\
\hline
MalConv       & 95.71 & 95.80 & 95.64 & 3.56 & 4.84 & 99.14 \\
\hline
\end{tabular}%
}
\end{table}

\section{Discussion}
\label{sec:discussion}
The results demonstrate that pre-training on byte-level data provides a strong 
initialization for downstream malware classification. The faster convergence of the 
fine-tuned model suggests that the pre-trained weights capture generalizable 
byte-level patterns that transfer effectively to executable files.

The initially higher training loss observed during the first epoch of fine-tuning is 
consistent with the semantic differences between image bytes and executable bytes. 
Despite this mismatch, the model rapidly adapts, achieving lower training loss than 
the scratch model by epoch two.

The pre-trained model was trained on byte sequences extracted from TIFF files. 
Although TIFF is an image format, it is also a structured binary container composed 
of headers, metadata, compressed blocks, and padding. Because 
\textbf{\mbox{MalwareFormer}} operates directly on raw bytes without visual inductive 
biases, it learns representations that capture the statistical and structural 
properties of binary streams rather than image semantics. These learned patterns 
likely facilitate transfer to executable analysis, helping explain the improved 
convergence and generalization behavior observed during fine-tuning.

To further characterize model behaviour, we stratify classification performance by 
input-window entropy, examining how byte randomness affects detection accuracy, FPR, 
and FNR across six entropy bins. As shown in Table~\ref{tab:entropy_analysis}, 
the model maintains balanced and stable error rates across the moderate entropy range 
(0--7), with both FPR and FNR remaining below 3.5\%. However, in the highest entropy 
bin (7--8), which contains 5,011 samples predominantly corresponding to packed or 
encrypted content, the model's FPR spikes to \textbf{14.24\%} while FNR drops to near zero (0.08\%). This reveals that the model 
has learned to strongly associate extreme entropy with malware, correctly capturing 
a common obfuscation pattern but at the cost of over-flagging high-entropy benign 
files such as compressed archives or resource-heavy executables. Notably, overall 
accuracy in this bin remains high (99.06\%), illustrating how aggregate accuracy 
can mask a severe class-asymmetric failure that would be operationally significant 
in deployment. The same asymmetry is observed when stratifying by full-file entropy 
(FPR: 11.63\%, FNR: 0.67\% in the 7--8 bin), confirming that this behaviour 
reflects a genuine bias in the learned representations rather than an artifact of 
input truncation. These findings suggest that entropy-aware training strategies or 
a dedicated high-entropy classifier could meaningfully reduce false positives in 
this regime.

\begin{table}[htbp]
\centering
\caption{Classification performance of \textbf{\mbox{MalwareFormer}} stratified by 
         input-window entropy bin. Accuracy, FPR, and FNR are reported per bin. 
         The 7--8 bin reveals a severe asymmetric failure: near-zero FNR but a 
         substantially elevated FPR, indicating systematic over-flagging of 
         high-entropy benign files.}
\label{tab:entropy_analysis}
\resizebox{\columnwidth}{!}{%
\begin{tabular}{|c|c|c|c|c|}
\hline
\textbf{Entropy bin} & \textbf{Samples} & \textbf{Accuracy (\%)} & 
\textbf{FPR (\%)} & \textbf{FNR (\%)} \\
\hline
0--2 & 86     & 98.84 & 0.00           & 1.82          \\
2--4 & 1,279  & 98.59 & 0.97           & 2.54          \\
4--5 & 3,165  & 98.64 & 1.17           & 1.54          \\
5--6 & 6,044  & 97.96 & 1.34           & 3.30          \\
6--7 & 14,648 & 98.18 & 2.07           & 1.62          \\
7--8 & 5,011  & 99.06 & \textbf{14.24} & \textbf{0.08} \\
\hline
\end{tabular}%
}
\end{table}

Comparison with CNN-based malware detectors highlights the advantages of 
transformer-based architectures for byte sequence modeling. CNNs use fixed receptive 
fields and local convolutional operations, which limit their ability to capture 
long-range dependencies. In contrast, \textbf{\mbox{MalwareFormer}} combines token 
reduction with windowed self-attention, enabling aggregation of both local and global 
byte-level structures. This architectural expressiveness likely contributes to its 
superior performance across all threshold-dependent metrics, even under truncated 
input constraints.

\section{Conclusion}
\label{sec:conclusion}
In this work, we presented \textbf{\mbox{MalwareFormer}}, a transformer-based, byte-level malware detection framework that operates directly on raw executable bytes without requiring handcrafted features or domain expertise. Leveraging transformer self-attention, the model captures long-range byte dependencies and consistently outperforms CNN-based baselines across all threshold-dependent metrics, achieving \textbf{98.35\%} accuracy, a macro-F1 of \textbf{98.32\%}, and a false positive rate of \textbf{1.94\%} on 50K-byte inputs.

We further analyzed the impact of input sequence length and found that while larger byte sequences can provide modest gains, the framework shows stable performance even with truncated inputs. An entropy-stratified analysis reveals a nuanced failure mode: while error rates remain balanced across moderate entropy ranges, the model exhibits an entropy-induced false positive bias in the highest entropy bin (7--8), near-zero FNR (0.08\%) but an elevated FPR (14.24\%), over-flagging compressed or encrypted benign files due to an overgeneralized association between extreme entropy and malware. Future work will explore entropy-aware training and hybrid approaches to 
address this bias.

Overall, \textbf{\mbox{MalwareFormer}} demonstrates that transformer architectures provide a powerful, scalable, and practical solution for byte-level malware analysis, eliminating the need for hand-crafted features.

\bibliography{references}
\bibliographystyle{IEEEtran}

\end{document}